\documentclass[reqno,12pt]{article}
\usepackage{amsfonts,amssymb,amsmath,amscd,bm,bbold,cite,delarray,subfigure}
\usepackage[dvips]{color}
\usepackage[dvips]{graphicx}

\numberwithin{equation}{section}

\font\capital=rsfs12
\font\scriptcapital=rsfs10 at 7 truept
\font\scriptscriptcapital=rsfs10 at 5 truept
\def\scri{\fam=15}
\newcommand{\mathscri}[1]{{{\scri #1}}}

\font\sansserif=cmss12
\font\scriptsansserif=cmss12 at 7 truept
\font\scriptscriptsansserif=cmss10 at 5 truept
\font\euler=eusm10 at 12 truept
\font\scripteuler=eusm7
\font\scriptscripteuler=eusm5 
\begin{document}

\hrule\vskip.4cm
\hbox to 14.5 truecm{August 2006\hfil DFUB 06--2}
\hbox to 14.5 truecm{Version 1  \hfil hep-th/0608145}
\vskip.4cm\hrule
\vskip.7cm
\begin{large}
\centerline{\bf THE BIHERMITIAN TOPOLOGICAL SIGMA MODEL}   
\end{large}
\vskip.2cm
\centerline{by}
\vskip.2cm
\centerline{\bf Roberto Zucchini}
\centerline{\it Dipartimento di Fisica, Universit\`a degli Studi di Bologna}
\centerline{\it V. Irnerio 46, I-40126 Bologna, Italy}
\centerline{\it I.N.F.N., sezione di Bologna, Italy}
\centerline{\it E--mail: zucchinir@bo.infn.it}
\vskip.7cm
\hrule
\vskip.7cm
\centerline{\bf Abstract} 
\par\noindent
BiHermitian geometry, discovered long ago by Gates, Hull and Ro\v cek, is
the most general sigma model target space geometry allowing 
for $(2,2)$ world sheet supersymmetry. 
By using the twisting procedure proposed by Kapustin and Li, we work out 
the type $A$ and $B$ topological sigma models for a general biHermtian 
target space, we write down the explicit expression of the sigma model's
action and BRST transformations and present a computation of the topological 
gauge fermion and the topological action.
\vskip.4cm
\par\noindent
Keywords: Supersymmetric Sigma Model, Topological Field Theory, Generalized Complex Geometry.

\vfill\eject

\begin{small}
\section{  \bf Introduction}
\label{sec:intro}
\end{small}

\vskip.3cm
Type II superstring Calabi--Yau compactifications are described by $(2,2)$ 
superconformal sigma models with Calabi--Yau target manifolds. These field theories
are however rather complicated and, so, they are difficult to study. 
In 1988, Witten showed that a $(2,2)$ supersymmetric sigma model on a 
Calabi--Yau space could be twisted in two different ways,   
to give the so called $A$ and $B$ topological sigma models \cite{Witten1,Witten2}.  
Unlike the original untwisted sigma model, the topological models are soluble:
the calculation of observables can be reduced to standard problems of geometry and topology.
For the $A$ model, the ring of observables is found to be a deformation of
the complex de Rham cohomology $\bigoplus_p H^p(M,\mathbb{C})_{\rm qu}$
going under the name of quantum cohomology. 
For the $B$ model, the ring of
observables turns out to be isomorphic to $\bigoplus_{p,q} H^p(\wedge^q T^{1,0} M)$. 
Furthermore, all correlators of the $A$ model are symplectic invariants of $M$, while all
correlators of the $B$ model are invariants of the complex structure on $M$.
For this reason, the topological sigma models constitute an ideal and
convenient field theoretic ground for the study of 2--dimensional
supersymmetric field theories. 

Witten's analysis was restricted to the case where the sigma model target 
space geometry was Kaehler. In a classic paper, Gates, Hull and Ro\v cek 
\cite{Gates} showed that, for a 2--dimensional sigma model, the most general
target space geometry allowing for $(2,2)$ supersymmetry 
was biHermitian or Kaehler with torsion geometry. This is characterized by 
a Riemannian metric $g_{ab}$, two generally non commuting complex structures 
$K_\pm{}^a{}_b$ and a closed $3$--form $H_{abc}$,
such that $g_{ab}$ is Hermitian with respect to both the $K_\pm{}^a{}_b$ and 
the $K_\pm{}^a{}_b$ are parallel with respect to two different metric
connections with torsion proportional to $\pm H_{abc}$ 
\cite{Rocek,Ivanov,Bogaerts,Lyakhovich}.
This geometry is more general than that considered by Witten, which
corresponds to the case where $K_+{}^a{}_b=\pm K_-{}^a{}_b$ and $H_{abc}=0$.
So, the natural question arises as to construct topological sigma models 
with biHermitian target space. 

A turning point in the quest towards accomplishing this goal was 
the realization that biHermitian geometry is naturally expressed in the language 
of generalized complex and Kaehler geometry
worked out by Hitchin and Gualtieri \cite{Hitchin1,Gualtieri,Zabzine3}. 
Many attempts have been made to construct sigma models with generalized 
complex or Kaehler target manifolds, by invoking world sheet supersymmetry, 
employing the Batalin--Vilkovisky quantization algorithm, etc.
\cite{Kapustin1,Kapustin2,Lindstrom2,Lindstrom3,
Lindstrom4,Lindstrom5,Chiantese,
Zabzine1,Zabzine2,Zucchini1,Zucchini2,Zucchini3,Li,
Pestun1}. All these were somehow unsatisfactory either because they
remained confined to the analysis of geometrical aspects of the sigma
models or because they yielded field theories, which though interesting in
their own, were not directly suitable for quantization, showed no apparent
kinship with Witten's $A$ and $B$ models and were of limited relevance for
string theory. 

In their seminal paper \cite{Kapustin2}, Kapustin and Li defined 
and studied the analogues of the $A$ and $B$ models for the 
general biHermitian $(2,2)$ supersymmetric sigma model. 
They tackled several crucial issues. 

\par\noindent
{\it a}) They formulated their analysis in the natural framework of 
generalized complex and Kaehler geometry. 

\par\noindent
{\it b}) They identified the appropriate twisting 
prescriptions yielding the biHermitian $A$ and $B$ models. 

\par\noindent
{\it c}) They showed that the consistency of the quantum theory requires 
one of the two twisted generalized complex structures forming 
the target space twisted generalized Kaehler structure to be a twisted generalized 
Calabi--Yau structure.

\par\noindent
{\it d}) They showed that the BRST cohomology is isomorphic to the  
cohomology of the Lie algebroid associated with that structure. 

\noindent 
However, Kapustin and Li left much work to be done. 

\par\noindent
{\it e}) They did not write down the explicit expression of the action $S_t$ of the 
biHermitian $A$ and $B$ models. 

\par\noindent
{\it f}) They provided only partial expressions of the BRST symmetry  
operator $s_t$.

\par\noindent
{\it g}) They left unsolved the problem of writing the action in the form 
\begin{equation}\label{0}
S_t=s_t\Psi_t+S_\mathrm{top},
\end{equation}
where $\Psi_t$ is a ghost number $-1$ gauge fermion and $S_\mathrm{top}$ is
a topological action, as required by the topological nature of the model. 

\par\noindent
In this paper, we have carried out these missing calculations and written down all 
the required expressions. It is our belief that the completeness of the theory 
definitely demands this work to be done. There still are open 
problems with point {\it g} above. Their solution is left for future work. 

The paper is organized as follows. In sect. \ref{sec:biHermitian},
we review the basic notions of biHermitian and generalized complex and Kaehler 
geometry used in the paper. In sect. \ref{sec:(2,2)susysigma}, 
we review the main properties of the biHermitian (2,2) supersymmetric 
sigma model, which are relevant in the following analysis.
In sect. \ref{sec:amodel}, we implement the twisting prescriptions of Kapustin
and Li and write down the explicit expressions of the action $S_t$ and of the
BRST symmetry operator $s_t$ of the 
biHermitian $A$ and $B$ models. In sect. \ref{sec:descent}, we study 
the ghost number anomaly and the descent formalism.
In sect. \ref{sec:gaugeferm}, we compute the gauge fermion  
$\Psi_t$ and the topological action $S_\mathrm{top}$ appearing in \eqref{0}.
Finally, in the appendices, we conveniently collect the technical
details of our analysis. 

After this work was completed, we became aware of the paper 
\cite{Chuang}, where similar results were obtained. 

\vfill\eject

\begin{small}
\section{  \bf BiHermitian geometry}
\label{sec:biHermitian}
\end{small}

The target space geometry of the sigma models studied in the
following is biHermitian. Below, we review the basic facts of 
biHermitian geometry and its relation to generalized Kaehler geometry.  

Let $M$ be a smooth manifold. A biHermitian structure $(g,H,K_\pm)$ on $M$ consists of
the following elements.
\par\noindent
$a$) A Riemannian metric $g_{ab}$ \footnote{\vphantom{$\Bigg]$} Here and below, indices are raised and lowered 
by using the metric $g_{ab}$.}.
\par\noindent
$b$) A closed 3--form $H_{abc}$  
\begin{equation}
\partial_{[a}H_{bcd]}=0.
\vphantom{\int}
\label{}
\end{equation}
\par\noindent
$c$) Two complex structures $K_\pm{}^a{}_b$,
\begin{align}
&K_\pm{}^a{}_cK_\pm{}^c{}_b=-\delta^a{}_b,
\vphantom{\int}
\label{}
\\
&K_\pm{}^d{}_a\partial_dK_\pm{}^c{}_b-K_\pm{}^d{}_b\partial_dK_\pm{}^c{}_a
-K_\pm{}^c{}_d\partial_aK_\pm{}^d{}_b+K_\pm{}^c{}_d\partial_bK_\pm{}^d{}_a=0.
\vphantom{\int}
\label{}
\end{align}
\par\noindent
They satisfy the following conditions. 
\par\noindent
$d$) $g_{ab}$ is Hermitian with respect to both $K_\pm{}^a{}_b$:
\begin{equation}
K_{\pm ab}+K_{\pm ba}=0.\vphantom{\int}
\label{}
\end{equation}
\par\noindent
$e$) 
The complex structures $K_\pm{}^a{}_b$ are parallel with respect to the
connections $\nabla_{\pm a}$ 
\begin{equation}
\nabla_{\pm a}K_\pm{}^b{}_c=0,
\label{}
\end{equation}
where the connection coefficients $\Gamma_\pm{}^a{}_{bc}$ are given by
\begin{equation}
\Gamma_\pm{}^a{}_{bc}=\Gamma^a{}_{bc}\pm\frac{1}{2}H^a{}_{bc}, 
\label{}
\end{equation}
$\Gamma^a{}_{bc}$ being the usual Levi--Civita connection coefficients.

The connections $\nabla_{\pm a}$ do have a non vanishing torsion
$T_{\pm abc}$, which is totally antisymmetric and indeed equal to the 
3--form $H_{abc}$ up to sign,
\begin{equation}
T_{\pm abc}=\pm H_{abc}.
\label{}
\end{equation}
The Riemann tensors $R_{\pm abcd}$ of the $\nabla_{\pm a}$ satisfy a number of
relations, the most relevant of which are collected in appendix \ref{sec:appendixtens}. 

Usually, in complex geometry, it is convenient to write the relevant
tensor identities in the complex coordinates of the underlying complex
structure rather than in general coordinates. 
In biHermitian geometry, one is dealing with two generally non
commuting complex structures. One could similarly write the 
tensor identities in the complex coordinates of either complex structures,
but, in this case, the convenience of complex versus general coordinates
would be limited. We decided, therefore, to opt for general coordinates 
throughout the paper. To this end, we define the complex tensors
\begin{equation}
\Lambda_{\pm}{}^a{}_b=\frac{1}{2}\big(\delta^a{}_b-iK_\pm{}^a{}_b\big).
\label{Pipm}
\end{equation}
The $\Lambda_{\pm}{}^a{}_b$ satisfy the relations 
\begin{subequations}\label{Pipm1}
\begin{align}
&\Lambda_{\pm}{}^a{}_c\Lambda_{\pm}{}^c{}_b=\Lambda_{\pm}{}^a{}_b,\vphantom{\int}
\label{Pipm1a}
\\
&\Lambda_{\pm}{}^a{}_b+\overline{\Lambda}{}_{\pm}{}^a{}_b=\delta^a{}_b,\vphantom{\int}
\label{Pipm1b}
\\
&\Lambda_{\pm}{}^a{}_b=\overline{\Lambda}{}_{\pm}{}_b{}^a.\vphantom{\int}
\label{Pipm1c}
\end{align}
\end{subequations}
Thus, the $\Lambda_{\pm}{}^a{}_b$ are projector valued endomorphisms of the
complexified tangent bundle $TM\otimes\mathbb{C}$.
The corresponding projection subbundles of $TM\otimes\mathbb{C}$ are the $\pm$ holomorphic
tangent bundles $T_\pm^{10}M$.

It turns out that the 3--form $H_{abc}$ is of type $(2,1)+(1,2)$ with respect
to both complex structures $K_\pm{}^a{}_b$,
\begin{equation}
H_{def}\Lambda_{\pm}{}^d{}_a\Lambda_{\pm}{}^e{}_b\Lambda_{\pm}{}^f{}_c=0
~~\text {and c.c.}
\label{}
\end{equation}
Other relations of the same type involving the Riemann tensors $R_{\pm abcd}$ are 
collected in appendix \ref{sec:appendixtens}. 

In \cite{Gualtieri}, Gualtieri has shown that biHermitian geometry is related 
to generalized Kaehler geometry. This, in turn, is part  
of generalized complex geometry. For a review of generalized complex and Kaehler 
geometry accessible to physicists, see \cite{Cavalcanti,Zabzine3}. Here, we shall restrict
ourselves to mention the salient points of these topics.

Let $H$ be a closed $3$--form. An $H$ twisted generalized complex structure 
$\mathcal{J}$ is a section of the endomorphism bundle of $TM\oplus T^*M$ such 
that $\mathcal{J}^2=-1$ and $\mathcal{J}=-\mathcal{J}^*$ with respect to 
the canonical inner product of $TM\oplus T^*M$ and $\mathcal{J}$ is 
integrable with respect to the $H$ twisted Courant brackets of 
$TM\oplus T^*M$.

There is a pure spinor formulation of generalized complex geometry, which is 
often very useful. Spinors of the Clifford bundle $C\ell(TM \oplus T^*M)$
are just sections of $\wedge^* T^*M$, i. e. non homogeneous forms. 
The Clifford action is defined by  
\begin{equation}\label{Cliffact}
(X+\xi) \cdot\phi=i_X\phi+\xi\wedge\phi,
\end{equation}
for $X+\xi$ a section of $TM \oplus T^*M$ and $\phi$ a section of $\wedge^* T^*M$.
With each nowhere vanishing spinor $\phi$, there is associated 
the subbundle $L_\phi$ of $TM \oplus T^*M$  
spanned by all sections $X+\xi$ of $TM \oplus T^*M$ such that 
\begin{equation}\label{Adotphi=0}
(X+\xi)\cdot\phi=0.
\end{equation}
$L_\phi$ is isotropic. 
The spinor $\phi$ is pure if $L_\phi$ is maximally isotropic. 
Conversely, with any maximally isotropic subbundle $L$ of $TM \oplus T^*M$, 
there is associated a nowhere vanishing pure spinor $\phi$ defined 
up to pointwise normalization such that $L=L_\phi$. In general, for a given $L$, 
$\phi$ is defined only locally. Thus, $L$ yields a generally non trivial
line bundle $U_L$ in $\wedge^* T^*M$, the pure spinor line of $L$. 
The above analysis continues to hold upon complexification. 

With any $H$ twisted generalized complex structure, there is associated 
a maximally isotropic subbundle $L_\mathcal{J}$ of 
$(TM \oplus T^*M)\otimes \mathbb{C}$: $L_\mathcal{J}$ is the $+i$
eigenbundle of $\mathcal{J}$ in $(TM \oplus T^*M)\otimes \mathbb{C}$. 
In turn, with $L_\mathcal{J}$, there is associated a pure spinor line $U_\mathcal{J}$ defined 
locally by a pure spinor $\phi_\mathcal{J}$.
The integrability of $\mathcal{J}$ is equivalent to 
\begin{equation}
d\phi_\mathcal{J}-H\wedge\phi_\mathcal{J}=(X+\xi)\cdot\phi_\mathcal{J},
\end{equation}
for some section $X+\xi$ of $(TM \oplus T^*M)\otimes \mathbb{C}$ \cite{Gualtieri}
\footnote{\vphantom{$\bigg[$} One further has the $\mathrm{Spin}_0(TM \oplus T^*M)$ invariant 
condition $[\phi_\mathcal{J}\wedge
\sigma(\bar\phi_\mathcal{J})]_\mathrm{top}\not=0$,
where $\sigma$ is the automorphism which reverses the order of the wedge
product and 
$[\cdots]_\mathrm{top}$ denotes projection on the top form.}.

An $H$ twisted generalized complex structure $\mathcal{J}$ is an $H$ twisted weak
generalized Calabi-Yau structure, if the nowhere vanishing pure spinor 
$\phi_\mathcal{J}$ is globally defined and further\hphantom{xxxxxxxxxxxxxx}
\begin{equation}\label{weakCY}
d\phi_\mathcal{J}-H\wedge \phi_\mathcal{J}=0. 
\end{equation}
Note that the line bundle $U_L$ is trivial in this case. 

If $\omega$ is a symplectic structure, then
\begin{equation}
\label{ASS}
\mathcal{J}_\omega=
\bigg(
\begin{array}{ll}
0 & -\omega^{-1} \\
\omega & ~~0
\end{array}
\bigg)
\end{equation}
is an untwisted generalized complex structure.
Its associated pure spinor is 
\begin{equation}
\phi_{\mathcal{J}_\omega}=\exp_\wedge(i \omega).
\end{equation}
$\phi_{\mathcal{J}_\omega}$ is globally defined and closed. Therefore, 
$\mathcal{J}_\omega$ is a weak generalized Calabi--Yau structure. 

If $K$ is a complex structure, then 
\begin{equation}
\label{ACS}
\mathcal{J}_K=\bigg(
\begin{array}{ll}
K & ~~0\\
0 & -K^t
\end{array}
\bigg)
\end{equation}
is an untwisted generalized complex structure. 
Its associated pure spinor is
\begin{equation}
\phi_{\mathcal{J}_K}=\Omega^{(n,0)} 
\end{equation}
where $\Omega^{(n,0)}$ is a closed holomorphic volume form.
$\phi_{\mathcal{J}_K}$ is only locally defined in general.
$\mathcal{J}_K$ is a weak generalized Calabi-Yau structure, if
$\Omega^{(n,0)}$ is globally defined. Note that this requires the
vanishing of the Chern class $c_1(M)$. 

An $H$ twisted generalized Kaehler structure structure
consists of a pair of $H$ twisted generalized complex structures
${\cal J}_1, {\cal J}_2$ such that 
${\cal J}_1, {\cal J}_2$ commute and 
${\cal G}\equiv -{\cal J}_1{\cal J}_2>0$
with respect to the canonical inner product of $TM\oplus T^*M$.

As shown in \cite{Gualtieri}, if $(g,H,K_\pm)$ is a biHermitian structure,
then 
\begin{equation}\label{J1/2}
 \mathcal{J}_{1/2} = \frac{1}{2}  
\bigg( \begin{array}{ll}
       K_+ \pm K_- & (K_+\mp K_-)g^{-1} \\
     g(K_+\mp K_-)& - (K_+{}^t \pm K_-{}^t)
\end{array} \bigg)
\end{equation}
yield an $H$-twisted generalized Kaehler structure as defined above. 

An ordinary Kaehler structure $(g,K)$ yields simultaneously a symplectic structure $\omega=gK$
and a complex structure $K$, with which there are associated the
generalized complex structures $\mathcal{J}_1=\mathcal{J}_K$ 
$\mathcal{J}_2=\mathcal{J}_\omega$ defined in \eqref{ASS}, \eqref{ACS}.
Then, $(\mathcal{J}_1, \mathcal{J}_2)$ is a generalized Kaehler structure.
If the Kaehler structure $(g,K)$ is Calabi--Yau, then both 
$\mathcal{J}_1$ $\mathcal{J}_2$ are weak generalized Calabi--Yau structures.

\vfill\eject

\begin{small}
\section{  \bf The (2,2) supersymmetric sigma model}
\label{sec:(2,2)susysigma}
\end{small}

We shall review next the main properties of the biHermitian (2,2) supersymmetric 
sigma model, which are relevant in the following analysis. 

The base space of the model is a $1+1$ dimensional Minkoskian surface
$\Sigma$, usually taken to be a cylinder.
The target space of the model is a smooth manifold $M$ equipped with a
biHermtian structure $(g,H,K_\pm)$. 
The basic fields of the model are the embedding field $x^a$ of $\Sigma$ into $M$ 
and the spinor fields $\psi_\pm{}^a$, which are valued in the vector bundle 
$x^*TM$ \footnote{\vphantom{$\bigg[$} Complying with an established use, 
here and in the following the indices $\pm$ are employed both to label the 
two complex structures $K_\pm$ of the relevant biHermitian structure and to denote
$2$--dimensional spinor indices. It should be clear from the context
what they stand for and no confusion should arise.}.

The action of biHermitian (2,2) supersymmetric 
sigma model is given by 
\begin{align}
S&=\int_\Sigma d^2\sigma\bigg[\frac{1}{2}(g_{ab}+b_{ab})(x)\partial_{++}x^a
\partial_{--}x^b
\label{22action}
\\
&\hskip1.9cm+\frac{i}{2}g_{ab}(x)(\psi_+{}^a\nabla_{+\,--}\psi_+{}^b
+\psi_-{}^a\nabla_{-\,++}\psi_-{}^b)
\nonumber\\
&\hskip1.9cm
+\frac{1}{4}R_{+abcd}(x)\psi_+{}^a\psi_+{}^b\psi_-{}^c\psi_-{}^d
\bigg],
\nonumber
\end{align}
where $\partial_{\pm\pm}=\partial_0\pm\partial_1$, \hphantom{xxxxxxxxxxx} 
\begin{equation}
\nabla_{\pm\,\mp\mp}=\partial_{\mp\mp}+\Gamma_\pm{}^\cdot{}_{c\,\cdot}(x)\partial_{\mp\mp}x^c
\end{equation}
and the field $b_{ab}$ is related to $H_{abc}$ as  
\begin{equation}
H_{abc}=\partial_ab_{bc}+\partial_bb_{ca}+\partial_cb_{ab}.
\end{equation}

The $(2,2)$ supersymmetry variations of the basic fields can be written in several ways. 
We shall write  them in the following convenient form
\begin{subequations}\label{22vars}
\begin{align}
\delta x^a&=
i\Big[\alpha^+\Lambda_+{}^a{}_b(x)\psi_+{}^b
+\tilde\alpha^+\overline{\Lambda}_+{}^a{}_b(x)\psi_+{}^b
\label{22varsa}\\
&\hphantom{=}~~~~+\alpha^-\Lambda_-{}^a{}_b(x)\psi_-{}^b
+\tilde\alpha^-\overline{\Lambda}_-{}^a{}_b(x)\psi_-{}^b\Big],~~~~~~~~~~~~~~~~~
\nonumber
\end{align}
\pagebreak[2]
\begin{align}
\delta\psi_\pm{}^a&=-\alpha^\pm\overline{\Lambda}_\pm{}^a{}_b(x)\partial_{\pm\pm}x^b
-\tilde\alpha^\pm\Lambda_\pm{}^a{}_b(x)\partial_{\pm\pm}x^b
\label{22varsb}\\
&\hphantom{=}\,-i\Gamma_\pm{}^a{}_{bc}(x)\Big[\alpha^+\Lambda_+{}^b{}_d(x)\psi_+{}^d
+\tilde\alpha^+\overline{\Lambda}_+{}^b{}_d(x)\psi_+{}^d
\nonumber\\
&\hphantom{\,=-i\Gamma_\pm{}^a{}_{bc}(x)\Big[}+\alpha^-\Lambda_-{}^b{}_d(x)\psi_-{}^d
+\tilde\alpha^-\overline{\Lambda}_-{}^b{}_d(x)\psi_-{}^d\Big]\psi_\pm{}^c
\nonumber\\
&\hphantom{=}\,\pm i H^a{}_{bc}(x)\Big[
\alpha^\pm\Lambda_\pm{}^b{}_d(x)\psi_\pm{}^d
+\tilde\alpha^\pm\overline{\Lambda}_\pm{}^b{}_d(x)\psi_\pm{}^d\Big]\psi_\pm{}^c
\nonumber\\
&\hphantom{=}\,\mp\frac{i}{2}
\big(
\alpha^\pm\Lambda_\pm{}^a{}_d
+\tilde\alpha^\pm\overline{\Lambda}_\pm{}^a{}_d\big)H^d{}_{bc}(x)
\psi_\pm{}^b\psi_\pm{}^c,
\nonumber
\end{align}
\end{subequations}
where $\alpha^\pm$, $\tilde\alpha^\pm$ are constant Grassmann parameters. 
$\delta$ generates a $(2,2)$ supersymmetry algebra on shell.
The action $S$ enjoys $(2,2)$ supersymmetry, so that
\begin{equation}
\delta S=0.
\end{equation}

The biHermitian $(2,2)$ supersymmetric sigma model is characterized also by 
two types of $R$ symmetry: the $U(1)_V$ vector $R$ symmetry
\begin{subequations}
\begin{align}
\delta_Vx^a&=0,
\\
\delta_V\psi_\pm{}^a&=-i\epsilon_V\Lambda_\pm{}^a{}_b(x)\psi_\pm{}^b
+i\epsilon_V\overline{\Lambda}_\pm{}^a{}_b(x)\psi_\pm{}^b,
\end{align}
\end{subequations}
and the $U(1)_A$ axial $R$ symmetry
\begin{subequations}
\begin{align}
\delta_Ax^a&=0,
\\
\delta_A\psi_\pm{}^a&=\mp i\epsilon_A\Lambda_\pm{}^a{}_b(x)\psi_\pm{}^b
\pm i\epsilon_A\overline{\Lambda}_\pm{}^a{}_b(x)\psi_\pm{}^b,
\end{align}
\end{subequations}
where $\epsilon_V$, $\epsilon_A$ are infinitesimal real parameters. 
Classically, the action $S$ enjoys both types of $R$ symmetry, so that 
\begin{equation}
\delta_V S=\delta_A S=0.
\end{equation}
As is well known, at the quantum level, the $R$ symmetries are spoiled by
anomalies in general. The $R$ symmetry anomalies
cancel, provided the following conditions are satisfied \cite{Kapustin2}:
\begin{subequations}\label{ancanc1}
\begin{align}
&c_1(T_+^{10}M)-c_1(T_-^{10}M)=0,~~~~~~~\text{vector $R$  symmetry},\\
&c_1(T_+^{10}M)+c_1(T_-^{10}M)=0,~~~~~~~\text{axial $R$  symmetry}.
\end{align}
\end{subequations} 

To generate topological sigma models using twisting, 
we switch to the Euclidean version of the $(2,2)$ supersymmetric 
sigma model. Henceforth, $\Sigma$ is a compact Riemann surface of genus 
$\ell_\Sigma$. Further, the following formal substitutions are to be 
implemented  
\begin{subequations}
\begin{align}
\partial_{++}&\rightarrow\partial_z
\\
\partial_{--}&\rightarrow\overline{\partial}_{\overline{z}}
\\
\psi_+{}^b&\rightarrow\psi_\theta{}^b\in C^\infty(\Sigma,\kappa_\Sigma{}^\frac{1}{2}\otimes x^*TM)
\\
\psi_-{}^b&\rightarrow\psi_{\overline{\theta}}{}^b\in C^\infty(\Sigma,
\overline{\kappa}_\Sigma{}^\frac{1}{2}\otimes x^*TM)
\end{align}
\end{subequations}
where $\kappa_\Sigma{}^\frac{1}{2}$ is any chosen spin structure
(a square root of the canonical line bundle $\kappa_\Sigma$ of $\Sigma$).

The topological twisting of the biHermitian $(2,2)$ supersymmetric sigma model 
is achieved by shifting the spin of fermions 
either by $q_V/2$ or $q_A/2$, where $q_V$, $q_A$ are the fermion's vector and axial
$R$ charges, respectively. The resulting topological sigma models will be
called biHermitian $A$ and $B$ models, respectively. The twisting can be
performed only if the corresponding $R$ symmetry is non anomalous, i.e 
if the conditions \eqref{ancanc1} are satisfied.
The \eqref{ancanc1} can be rephrased as 
conditions on the the generalized Kaehler structure $(\mathcal{J}_1,\mathcal{J}_2)$ 
corresponding to the given biHermitian structure $(g,H,K_\pm)$
according to \eqref{J1/2} \cite{Kapustin2}. If $E_k$ denotes the $+i$ eigenbundle of 
$\mathcal{J}_k$ in $(TM \oplus T^*M)\otimes \mathbb{C}$, then the conditions read
\begin{subequations}\label{ancanc2}
\begin{align}
&c_1(E_2)=0,~~~~~~~~~\text{$A$ twist},\\
&c_1(E_1)=0,~~~~~~~~~\text{$B$ twist}.
\end{align}
\end{subequations}

$R$ symmetry anomaly cancellation, however, is not sufficient by itself 
to ensure the consistency of the
twisting. Requiring the nilpotence of the BRST charge implies further conditions,
namely that
\begin{subequations}\label{brstancanc}
\begin{align}
&d\phi_2-H\wedge\phi_2=0,~~~~~~~~~\text{$A$ twist},\\
&d\phi_1-H\wedge\phi_1=0,~~~~~~~~~\text{$B$ twist},
\end{align}
\end{subequations}
where the $\phi_k$ are the globally defined pure spinors associated with the 
generalized complex structures $\mathcal{J}_k$ \cite{Kapustin2}. 
 
The conditions \eqref{ancanc2}, \eqref{brstancanc} 
are satisfied if the structures $\mathcal{J}_2$, $\mathcal{J}_1$ 
are twisted weak generalized Calabi--Yau, for the $A$ and $B$ twist, respectively. 
Further, when this is the case, the BRST cohomology is equivalent to the Lie
algebroid cohomology of the relevant generalized complex structure \cite{Kapustin2}.
This remarkable result was one of the achievements of Kapustin's and Li's work.

\vfill\eject

\begin{small}
\section{  \bf The biHermitian $A$ and $B$ sigma models}
\label{sec:amodel}
\end{small}

As explained in sect. \ref{sec:(2,2)susysigma},
the biHermitian $A$ and $B$ sigma models are obtained from the biHermitian 
$(2,2)$ supersymmetric sigma model via a set of formal prescriptions, 
called $A$ and $B$ twist. Concretely, the field content of the biHermitian $A$ sigma model is obtained 
from that of the $(2,2)$ supersymmetric sigma model via the substitutions
\begin{subequations}\label{Atwist}
\begin{align}
\Lambda_+{}^a{}_b(x)\psi_\theta{}^b&\rightarrow \chi_+{}^a\in C^\infty(\Sigma,x^*T^{10}_+M),
\\
\overline{\Lambda}_+{}^a{}_b(x)\psi_\theta{}^b&\rightarrow
\overline{\psi}_{+z}{}^a\in C^\infty(\Sigma,\kappa_\Sigma\otimes x^*T^{01}_+M),
\\
\Lambda_-{}^a{}_b(x)\psi_{\overline{\theta}}{}^b&\rightarrow
\psi_{-\overline{z}}{}^a \in C^\infty(\Sigma,\overline{\kappa}_\Sigma\otimes x^*T^{10}_-M),
\\
\overline{\Lambda}_-{}^a{}_b(x)\psi_{\overline{\theta}}{}^b&\rightarrow
\overline{\chi}_-{}_a\in C^\infty(\Sigma,x^*T^{01}_-M).
\end{align}
\end{subequations}
The symmetry variations of the $A$ sigma model fields are obtained from those 
of the $(2,2)$ supersymmetric sigma model fields  (cf. eq. \eqref{22vars}), by setting
\begin{subequations}\label{Atwistvars}
\begin{align}
\tilde\alpha^+=\alpha^-=0,
\\
\alpha^+=\tilde\alpha^-=\alpha.
\end{align}
\end{subequations}
Similarly, the field content of the biHermitian $B$ sigma model is obtained 
from that of the $(2,2)$ supersymmetric sigma model via the substitutions
\begin{subequations}\label{Btwist}
\begin{align}
\Lambda_+{}^a{}_b(x)\psi_\theta{}^b&\rightarrow 
\psi_{+z}{}^a\in C^\infty(\Sigma,\kappa_\Sigma\otimes x^*T^{10}_+M),
\\
\overline{\Lambda}_+{}^a{}_b(x)\psi_\theta{}^b&\rightarrow
\overline{\chi}_+{}^a\in C^\infty(\Sigma,x^*T^{01}_+M),
\\
\Lambda_-{}^a{}_b(x)\psi_{\overline{\theta}}{}^b&\rightarrow
\psi_{-\overline{z}}{}^a \in C^\infty(\Sigma,\overline{\kappa}_\Sigma\otimes x^*T^{10}_-M),
\\
\overline{\Lambda}_-{}^a{}_b(x)\psi_{\overline{\theta}}{}^b&\rightarrow
\overline{\chi}_-{}_a\in C^\infty(\Sigma,x^*T^{01}_-M).
\end{align}
\end{subequations}
The symmetry variations of the $B$ sigma model fields are obtained from those 
of the $(2,2)$ supersymmetric sigma model fields, by setting
\begin{subequations}\label{Btwistvars}
\begin{align}
\alpha^+=\alpha^-=0,
\\
\tilde\alpha^+=\tilde\alpha^-=\alpha.
\end{align}
\end{subequations}
Inspection of the $A$, $B$ twist prescriptions reveals that
\begin{equation}\label{ABcomp}
\text{$A$ twist}~\leftrightarrows~\text{$B$ twist}  
~~\text{under} ~~K_+{}^a{}_b  \leftrightarrows -K_+ {}^a{}_b.
\end{equation}
The target space geometrical data $(g,H,K_\pm)$, 
$(g,H_,\mp K_\pm)$ have precisely the same properties: they are both biHermitian
structures. So, at the classical level,
any statement concerning the $A$ ($B$)
model translates automatically into one concerning the $B$ ($A$) model
upon reversing the sign of $K_+$ 
\footnote{\vphantom{$\bigg[$} For notational consistency, 
exchanging $K_+{}^a{}_b  \leftrightarrows -K_+ {}^a{}_b$ must be accompanied by 
switching $\alpha^+\leftrightarrows \tilde\alpha^+$.}. For this reason, below,
we shall consider only the $B$ twist, unless otherwise stated. 

The twisted action $S_t$ is obtained from the $(2,2)$ supersymmetric sigma
model action $S$ \eqref{22action} implementing the substitutions
\eqref{Btwist}. One finds
\begin{align}
S_t&=\int_\Sigma d^2z\bigg[\frac{1}{2}(g_{ab}+b_{ab})(x)\partial_zx^a
\overline{\partial}_{\overline{z}}x^b
\label{topaction}\\
&\hskip1.9cm+ig_{ab}(x)(\psi_{+z}{}^a\overline{\nabla}_{+\overline{z}}\overline{\chi}_+{}^b
+\psi_{-\overline{z}}{}^a\nabla_{-z}\overline{\chi}{}_-{}^b)
\nonumber\\
&\hskip1.9cm
+R_{+abcd}(x)\overline{\chi}_+{}^a\psi_{+z}{}^b\overline{\chi}{}_-{}^c\psi_{-\overline{z}}{}^d
\bigg].
\nonumber
\end{align}
Similarly the twisted field variations are obtained from the $(2,2)$
supersymmetry field variations \eqref{22vars} via \eqref{Btwist}, \eqref{Btwistvars}.
One finds that 
\begin{equation}\label{deltatop}
\delta_t=\alpha s_t,
\end{equation}
where $s_t$ is the fermionic variation operator defined by
\begin{subequations}\label{stop}
\begin{align}
s_tx^a&=i(\overline{\chi}_+{}^a+\overline{\chi}{}_-{}^a),
\\
s_t\overline{\chi}_+{}^a&=-i\Gamma_+{}^a{}_{cb}(x)\overline{\chi}{}_-{}^c\overline{\chi}_+{}^b,
\\
s_t\overline{\chi}_-{}^a&=-i\Gamma_-{}^a{}_{cb}(x)\overline{\chi}_+{}^c\overline{\chi}{}_-{}^b,
\nonumber\\
s_t\psi_{+z}{}^a&=
-i\Gamma_+{}^a{}_{cb}(x)(\overline{\chi}_+{}^c+\overline{\chi}{}_-{}^c)\psi_{+z}{}^b
\\
&\hskip3.5cm-\Lambda_+{}^a{}_b(x)(\partial_zx^b
-iH^b{}_{cd}(x)\overline{\chi}_+{}^c\psi_{+z}{}^d),
\nonumber\\
s_t\psi_{-\overline{z}}{}^a&=
-i\Gamma_-{}^a{}_{cb}(x)(\overline{\chi}_+{}^c+\overline{\chi}{}_-{}^c)\psi_{-\overline{z}}{}^b
\nonumber\\
&\hskip3.5cm-\Lambda_-{}^a{}_b(x)(\overline{\partial}_{\overline{z}}x^b
+iH^b{}_{cd}(x)\overline{\chi}{}_-{}^c\psi_{-\overline{z}}{}^d).
\nonumber
\end{align}
\end{subequations}
The action $S_t$ is invariant under $s_t$,
\begin{equation}\label{stopSt=0}
s_tS_t=0.
\end{equation}
It is straightforward to verify that
the ideal of field equations in the algebra of local composite fields is
invariant under $s_t$. One verifies also that \hphantom{xxxxxxxxxxxxxxxxxx}
\begin{equation}
s_t{}^2\approx 0,
\end{equation}
where $\approx$ denotes equality on shell, so that $s_t$ is nilpotent on shell.
The proof of these statements is outlined in appendix \ref{sec:appendixsfeq}.
In this way, $s_t$ defines an on shell
cohomological complex. $s_t$ corresponds to the BRST charge of the model
and its on shell cohomology is isomorphic to the BRST cohomology.

In \eqref{Btwistvars}, there is no real need for the supersymmetry parameters
$\tilde\alpha^+$, $\tilde\alpha^-$ to take the same value $\alpha$, since,
under twisting both become scalars. If we insist $\tilde\alpha^+$,
$\tilde\alpha^-$ to be independent in \eqref{22vars}, we obtain a more general symmetry
variation
\begin{equation}\label{hatdeltatop}
\hat{\delta}_t=\tilde\alpha^+s_{t+}+\tilde\alpha^-s_{t+}
\end{equation}
where the fermionic variation operators $s_{t\pm}$ are given by 
\begin{subequations}\label{stoppm}
\begin{align}
s_{t+}x^a&=i\overline{\chi}_+{}^a,
\\
s_{t-}x^a&=i\overline{\chi}{}_-{}^a,
\nonumber\\
s_{\vphantom{f}t+}\overline{\chi}_+{}^a&=0,
\\
s_{\vphantom{f}t-}\overline{\chi}_+{}^a&=-i\Gamma_+{}^a{}_{cb}(x)\overline{\chi}{}_-{}^c\overline{\chi}_+{}^b,
\nonumber\\
s_{t+}\overline{\chi}_-{}^a&=-i\Gamma_-{}^a{}_{cb}(x)\overline{\chi}_+{}^c\overline{\chi}{}_-{}^b,
\nonumber\\
s_{t-}\overline{\chi}_-{}^a&=0,
\nonumber\\
s_{t+}\psi_{+z}{}^a&=
-i\Gamma_+{}^a{}_{cb}(x)\overline{\chi}_+{}^c\psi_{+z}{}^b
-\Lambda_+{}^a{}_b(x)(\partial_zx^b
-iH^b{}_{cd}(x)\overline{\chi}_+{}^c\psi_{+z}{}^d),
\\
s_{t-}\psi_{+z}{}^a&=-i\Gamma_+{}^a{}_{cb}(x)\overline{\chi}_-{}^c\psi_{+z}{}^b,
\nonumber\\
s_{t+}\psi_{-\overline{z}}{}^a&=
-i\Gamma_-{}^a{}_{cb}(x)\overline{\chi}_+{}^c\psi_{-\overline{z}}{}^b,
\nonumber\\
s_{t-}\psi_{-\overline{z}}{}^a&=
-i\Gamma_-{}^a{}_{cb}(x)\overline{\chi}{}_-{}^c\psi_{-\overline{z}}{}^b
-\Lambda_-{}^a{}_b(x)(\overline{\partial}_{\overline{z}}x^b
+iH^b{}_{cd}(x)\overline{\chi}{}_-{}^c\psi_{-\overline{z}}{}^d).
\nonumber
\end{align}
\end{subequations}
The action $S_t$ is invariant under both $s_{t\pm}$,
\begin{equation}\label{stoppmSt=0}
s_{t\pm}S_t=0.
\end{equation}
It is straightforward though lengthy to verify that
the ideal of field equations in the algebra of local composite fields is
invariant under each $s_{t\pm}$ separately. One can show also that the 
$s_{t\pm}$ are nilpotent and anticommute on shell
\begin{subequations}\label{stoppm2=0}
\begin{align}
&s_{t\pm}{}^2\approx 0,
\\
&s_{t+}s_{t-}+s_{t-}s_{t+}\approx 0.
\end{align}
\end{subequations}
The proof of these relations is outlined again in appendix \ref{sec:appendixsfeq}.
It is easy to verify that $s_t$ and the $s_{t\pm}$  are related as \hphantom{xxxxxxxxxxxxxxxx}
\begin{equation}\label{st=st++st-}
s_t=s_{t+}+s_{t-}.
\end{equation}
Therefore,  the $s_{t\pm}$ define an on shell
cohomological double complex, whose total differential is $s_t$, a fact 
already noticed in \cite{Kapustin2}. \eqref{st=st++st-} corresponds to the
decomposition of BRST charge in its left and right chiral components. 

The significance of these findings in not clear to us, beyond 
their ostensible algebraic meaning. 
As shown in \cite{Kapustin2}, the on shell $s_t$ cohomology, or BRST
cohomology, is equivalent to the Lie algebroid cohomology of the $H$ twisted generalized 
complex structure $\mathcal{J}_1$
underlying the target space 
biHermitian structure. No interpretation of the double on shell $s_{t\pm}$ cohomology 
on the same lines is known to us yet. 

With each biHermitian sigma model of the form described above, there is 
associated in canonical fashion a conjugate biHermitian sigma model as follows.
If $(g,H,K_\pm)$ is the target space biHermitian structure
of the given sigma model, the biHermitian structure
$(g',H',K')$ of the conjugate model is given by 
\begin{subequations}\label{target'}
\begin{align}
&g'{}_{ab}=g_{ab},
\\
&H'{}_{abc}=-H_{abc},
\\
&K'{}_\pm{}^a{}_b=K_\mp{}^a{}_b.
\end{align}
\end{subequations} 
The world sheet complex structure of the conjugate model is the conjugate 
of the world sheet complex structure of the given model.
The fields of the conjugate model are related to fields of the given model as
\begin{subequations}\label{fields'}
\begin{align}
&x'^a=x^a,
\\
&\overline{\chi}{}'{}_+{}^a=\overline{\chi}{}_-{}^a, 
\qquad \overline{\chi}{}'{}_-{}^a=\overline{\chi}{}_+{}^a, 
\\
&\psi'{}_{+z'}{}^a=\psi_{-\overline{z}}{}^a, 
\qquad\psi'{}_{-\overline{z}'}{}^a=\psi_{+z}{}^a,
\end{align}
\end{subequations} 
where $z'=\overline{z}$.
It is readily verified that the actions of the given and conjugate model are
equal \hphantom{xxxxxxxxxxxxxxxxxxxxx}
\begin{equation}
S'{}_t=S_t.
\end{equation}
Their BRST variations are likewise equal,
\begin{equation}
s'{}_t=s_t.
\end{equation}
Explicitly, this relations means that for any sigma model field $\phi$,
$s'{}_t\phi'=s_t\phi$ upon taking \eqref{target'},
\eqref{fields'} into account. Similarly, one has
\begin{equation}
s'{}_{t\pm}=s_{t\mp}.
\end{equation}

The original Kaehler $B$ model studied by Witten in \cite {Witten1,Witten2} 
is a particular case of the biHermitian $B$ model expounded here: 
the Kaehler $B$ model with target space Kaehler structure $(g,K)$ 
is equal to the biHermitian $B$ model with target space 
biHermitian structure $(g,H=0,K_\pm=K)$ up to simple field redefinitions. 
Similarly, the Kaehler $A$ model with Kaehler structure $(g,K)$
equals the biHermitian $A$ model with biHermitian structure $(g,H=0,K_\pm=\mp K)$.

\vfill\eject

\begin{small}
\section{  \bf Ghost number and descent}
\label{sec:descent}
\end{small}

We shall postpone the analysis of the delicate issue whether the biHermitian sigma
models described in sect. \ref{sec:amodel} are indeed topological field
theories to sect. \ref{sec:gaugeferm}. In this section, we shall study certain
properties of the models which are relevant in the computation of topological correlators,
namely the ghost number anomaly and the descent formalism. For reasons
explained in sect. \ref{sec:amodel}, we can restrict ourselves to the analysis
of the $B$ model. 

The biHermitian action $S_t$, given in eq. \eqref{topaction}, enjoys, besides
the BRST symmetry, the ghost number symmetry, defined by the field variations 
\begin{subequations}
\begin{align}
\delta_\mathrm{gh}x^a&=0,
\\
\delta_\mathrm{gh}\overline{\chi}_+{}^a&=-i\epsilon_+\overline{\chi}_+{}^a,
\\
\delta_\mathrm{gh}\overline{\chi}_-{}^a&=-i\epsilon_-\overline{\chi}_-{}^a,
\nonumber\\
\delta_\mathrm{gh}\psi_{+z}{}^a&=i\epsilon_+\psi_{+z}{}^a,
\\
\delta_\mathrm{gh}\psi_{-\overline{z}}{}^a&=i\epsilon_-\psi_{-\overline{z}}{}^a,
\nonumber
\end{align}
\end{subequations}
where $\epsilon_\pm$ are infinitesimal even parameters. Thus, 
\begin{equation}
\delta_\mathrm{gh}S_t=0.
\end{equation}
The fields $x^a$, $\overline{\chi}_+{}^a$, $\overline{\chi}_-{}^a$,
$\psi_{+z}{}^a$, $\psi_{-\overline{z}}{}^a$ have ghost number $0$, $+1$, $+1$, 
$-1$, $-1$, respectively. 
The fermionic variation operators $s_t$ or $s_{t\pm}$ all carry ghost number $+1$: their
action increases ghost number by one unit. 

At the quantum level, the ghost number symmetry is anomalous. Indeed, 
inspecting the fermionic kinetic terms of the action $S_t$, through a simple
application of the index theorem, it is easy to see that
\begin{subequations}
\begin{align}
n(\overline{\chi}_+)-n(\psi_{+z})
&=\int_\Sigma x^*c_1(T_+^{10}M)+\dim_\mathbb{C}M(1-\ell_\Sigma),
\\
n(\overline{\chi}_-)-n(\psi_{-\overline{z}})
&=\int_\Sigma x^*c_1(T_-^{10}M)+\dim_\mathbb{C}M(1-\ell_\Sigma),
\end{align}
\end{subequations}
where $n(\overline{\chi}_+)$, $n(\psi_{+z})$
$n(\overline{\chi}_-)$, $n(\psi_{-\overline{z}})$,
are the numbers of $\overline{\chi}_+{}^a$, $\psi_{+z}{}^a$,
$\overline{\chi}_-{}^a$, $\psi_{-\overline{z}}{}^a$ 
zero modes, respectively. Generically, $n(\psi_{+z})$,
$n(\psi_{-\overline{z}})$ vanish, while $n(\overline{\chi}_+)$, $n(\overline{\chi}_-)$
do not. Consequently, the vacuum carries a non vanishing ghost number charge
signaling an anomaly. In quantum correlators, this charge must be soaked up 
by insertions of fields $\overline{\chi}_+{}^a$, $\overline{\chi}_-{}^a$. 

Next, let us consider the field variations
corresponding to the symmetry parameters $\alpha^+$, $\alpha^-$
in \eqref{22vars}. 
This means that, in \eqref{Btwistvars}, we relax the condition 
$\alpha^+=\alpha^-=0$. Upon twisting, $\alpha^+$, $\alpha^-$ 
become Grassmann world sheet vector fields $\alpha^z$, $\alpha^{\overline{z}}$,
respectively. Thus, the corresponding fermionic variation operators $h_{t+z}$,
$h_{t-\overline{z}}$ are not scalar: they change the world sheet covariance 
properties of the fields as indicated by their notation. 
From \eqref{22vars}, we obtain easily 
\begin{subequations}\label{htoppm}
\begin{align}
h_{t+z}x^a&=i\psi_{+z}{}^a,
\\
h_{t-\overline{z}}x^a&=i\psi_{-\overline{z}}{}^a,
\nonumber\\
h_{t+z}\overline{\chi}_+{}^a&=-i\Gamma_+{}^a{}_{bc}(x)\psi_{+z}{}^b\overline{\chi}_+{}^c
-\overline{\Lambda}_+{}^a{}_b(x)(\partial_zx^b
-iH^b{}_{cd}(x)\psi_{+z}{}^c\overline{\chi}_+{}^d),
\\
h_{t-\overline{z}}\overline{\chi}_+{}^a
&=-i\Gamma_+{}^a{}_{bc}(x)\psi_{-\overline{z}}{}^b\overline{\chi}_+{}^c,
\nonumber\\
h_{t+z}\overline{\chi}_-{}^a&=-i\Gamma_-{}^a{}_{bc}(x)\psi_{+z}{}^b\overline{\chi}_-{}^c,
\nonumber\\
h_{t-\overline{z}}\overline{\chi}_-{}^a
&=-i\Gamma_-{}^a{}_{bc}(x)\psi_{-\overline{z}}{}^b\overline{\chi}{}_-{}^c
-\overline{\Lambda}_-{}^a{}_b(x)(\overline{\partial}_{\overline{z}}x^b
+iH^b{}_{cd}(x)\psi_{-\overline{z}}{}^c\overline{\chi}{}_-{}^d),
\nonumber
\\
h_{t+z}\psi_{+z}{}^a&=-i\Gamma_-{}^a{}_{bc}(x)\psi_{+z}{}^b\psi_{+z}{}^c,
\\
h_{t-\overline{z}}\psi_{+z}{}^a&=-i\Gamma_+{}^a{}_{bc}(x)\psi_{-\overline{z}}{}^b\psi_{+z}{}^c,
\nonumber
\\
h_{t+z}\psi_{-\overline{z}}{}^a&=-i\Gamma_-{}^a{}_{bc}(x)\psi_{+z}{}^b\psi_{-\overline{z}}{}^c,
\nonumber
\\
h_{t-\overline{z}}\psi_{-\overline{z}}{}^a&=
-i\Gamma_+{}^a{}_{bc}(x)\psi_{-\overline{z}}{}^b\psi_{-\overline{z}}{}^c.
\nonumber
\end{align}
\end{subequations}
The variation operators $h_{t+z}$, $h_{t-\overline{z}}$ lead to no new symmetry of the
action $S_t$. They would, if the world sheet vector fields $\alpha^z$, $\alpha^{\overline{z}}$
could be taken (anti)holomorphic, but this is not possible on a generic
compact Riemann surface $\Sigma$.
However, they are useful, as they implement the descent sequence 
yielding the world sheet $1$-- and $2$--form
descendants $\mathcal{O}^{(1)}$, $\mathcal{O}^{(2)}$ 
of an $s_t$ invariant world sheet $0$--form field $\mathcal{O}^{(0)}$ \cite{Witten1,Witten2}.  
Let us recall briefly how this works out in detail. 

Define the $1$--form bosonic variation operators
\footnote{\vphantom{$\Bigg[$} We assume conventionally that the $dz$,
$d\overline{z}$ anticommute with the fermionic fields $\overline{\chi}_\pm{}^a$, 
$\psi_{+z}{}^a$, $\psi_{+\overline{z}}{}^a$
and the fermionic variation operartors $s_t$, $s_{t\pm}$, $h_{t+z}$, $h_{t-\overline{z}}$.} 
\begin{equation}\label{htoppm1f}
h_{t+}=dzh_{t+z},\qquad h_{t-}=d\overline{z}h_{t-\overline{z}}
\end{equation}
acting on the algebra of form fields generated by the
fields $x^a$, $\overline{\chi}_+{}^a$, $\overline{\chi}_-{}^a$ and
the bosonic world sheet $1$--form fields 
\begin{equation}\label{formpsi}
\psi_+{}^a=dz\psi_{+z}{}^a,\qquad \psi_-{}^a=d\overline{z}\psi_{+\overline{z}}{}^a.
\end{equation}
Now, set \hphantom{xxxxxxxxxxxxxxxxxxxxxxxxxxx}
\begin{equation}\label{ht}
h_t=h_{t+}+h_{t-}.
\end{equation}
It is straightforward to verify that
the ideal of field equations in the algebra of local composite form fields is
invariant under $h_t$ and that the on shell relation
\begin{equation}\label{hstfos}
h_ts_t-s_th_t\approx-id,
\end{equation}
holds, where $d=dz\partial_z+d\overline{z}\partial_{\overline{z}}$ is the world sheet
de Rham differential. The proof of these results is outlined again in appendix
\ref{sec:appendixsfeq}.

Assume now that $\mathcal{O}^{(0)}$ is local $0$--form field such that \hphantom{xxxxxxxxxxxxxxxxx}
\begin{equation}\label{sO0=0}
s_t\mathcal{O}^{(0)}\approx 0.
\end{equation}
Define the $1$-- and $2$--form local fields 
\begin{subequations}\label{descent}
\begin{align}
\mathcal{O}^{(1)}&=h_t\mathcal{O}^{(0)},
\\
\mathcal{O}^{(2)}&=\frac{1}{2}h_t\mathcal{O}^{(1)}.
\end{align}
\end{subequations}
Then, from \eqref{hstfos}, \eqref{sO0=0},  one has the descent equations
\begin{subequations}\label{descenteq}
\begin{align}
s_t\mathcal{O}^{(1)}&\approx id\mathcal{O}^{(0)},
\\
s_t\mathcal{O}^{(2)}&\approx id\mathcal{O}^{(1)}.
\end{align}
\end{subequations}
Consequently, one has
\begin{subequations}\label{nonlocobs}
\begin{align}
s_t\oint_\gamma\mathcal{O}^{(1)}&\approx 0,
\\
s_t\oint_\Sigma\mathcal{O}^{(2)}&\approx 0,
\end{align}
\end{subequations}
where $\gamma$ is a $1$--cycle in $\Sigma$. In this way, non local BRST
invariants can be obtained canonically once a local scalar one is given. 
These invariants are the operators inserted in topological correlators
of the associated topological field theories.

The action of the $h_{t\pm}$ is in fact compatible with the double on shell $s_{t\pm}$ 
cohomology underlying the on shell $s_t$ cohomology. Indeed 
the ideal of field equations in the algebra of form fields is separately 
invariant under the $h_{t\pm}$ and, furthermore, the on shell relations
\begin{subequations}\label{hspm1fos}
\begin{align}
&h_{t+}s_{t+}-s_{t+}h_{t+}\approx-i\partial,
\\
&h_{t-}s_{t-}-s_{t-}h_{t-}\approx-i\overline{\partial},
\\
&h_{t+}s_{t-}-s_{t-}h_{t+}\approx 0,
\\
&h_{t-}s_{t+}-s_{t+}h_{t-}\approx 0
\end{align}
\end{subequations}
hold, where $\partial=dz\partial_z$ and c.c. are the world sheet Dolbeault
operators. One has further on shell relations
\begin{subequations}\label{hhpm1fos}
\begin{align}
&h_{t\pm}h_{t\pm}\approx 0,
\\
&h_{t+}h_{t-}-h_{t-}h_{t+}\approx 0.
\end{align}
\end{subequations} 
See again appendix \ref{sec:appendixsfeq} for a proof of these relations.

We note that the operators $h_t$, $h_{t\pm}$ all carry ghost number $-1$. 
Under conjugation (cf. eq. \eqref{target'}, \eqref{fields'}), one has 
\begin{equation}
h'{}_t=h_t
\end{equation}
and \hphantom{xxxxxxxxxxxxxxxxx}
\begin{equation}
h'{}_{t\pm}=h_{t\mp}.
\end{equation}

\vfill\eject

\begin{small}
\section{  \bf The biHermitian models are topological}
\label{sec:gaugeferm}
\end{small}
The biHermitian sigma models studied in sect. \ref{sec:amodel} should be
topological field theories. To check this, one should be able to express 
the sigma model action as
\begin{equation}\label{St=deltaPsi}
S_t\approx s_t\Psi_t+S_\mathrm{top},
\end{equation}
where $\approx$ denotes on shell equality, 
$\Psi_t$ is a ghost number $-1$ topological gauge fermion and $S_\mathrm{top}$ is
a topological action. General arguments indicate that, at the quantum level, when
\eqref{St=deltaPsi} holds,   
the topological sigma model field theory depends generically only on the
geometrical data contained in $S_\mathrm{top}$, since variations of the
geometrical data contained in $\Psi_t$ result in the insertion
in topological correlators of BRST cohomologically trivial operators and, so, cannot modify
those correlators \cite{Witten1,Witten2}. For reasons explained in
sect. \ref{sec:amodel}, below we shall restrict ourselves 
to the analysis of the $B$ model. 

In general, the topological action $S_\mathrm{top}$ is of the form 
\begin{equation}\label{Stopprop}
S_\mathrm{top}=\int_\Sigma x^*\omega,
\end{equation}
where $\omega$ is a $2$--form depending on some combinations of the target
space geometrical data $(g,H,K_\pm)$. If \eqref{St=deltaPsi}, \eqref{Stopprop}
hold, the sigma model field theory depends only on those combinations and is 
independent from the complex structure of the world sheet $\Sigma$. 
If $\omega$ is closed, \hphantom{xxxxxxxxxxxxxxxxxx}
 \hphantom{xxxxxxxxxxxxxx}
\begin{equation}\label{domega=0}
d\omega=0,
\end{equation}
then $S_\mathrm{top}$ is invariant under arbitrary infinitesimal variations of $x$. 
This condition, however, is not strictly necessary to show the topological nature of
the model, though it holds normally 
\footnote{\vphantom{$\Bigg]$} We thank A. Kapustin for pointing this out to us.}. 
When \eqref{domega=0} holds, we say that $S_\mathrm{top}$ is strictly topological. 

When $H=0$, it is straightforward to see that $\Psi_t$, $S_{\mathrm{top}}$ are given by 
\begin{subequations}\label{PsiStopH=0}
\begin{align}
\Psi_t&=-\int_\Sigma d^2z\frac{1}{2}g_{ab}(x)
\big(\psi_{+z}{}^a\overline{\partial}_{\overline{z}}x^b+\psi_{-\overline{z}}{}^a\partial_zx^b\big),
\label{PsiH=0}\\
S_\mathrm{top}&=\int_\Sigma d^2z\frac{1}{4}\big(2b_{ab}
-iK_{+ab}+iK_{-ab}\big)(x)\partial_zx^a\overline{\partial}_{\overline{z}}x^b.
\label{StopH=0}
\end{align}
\end{subequations}
The expression of $\Psi_t$ is formally identical to that originally found by
Witten in \cite{Witten1,Witten 2}. The action $S_\mathrm{top}$ is of the form
\eqref{Stopprop}, \eqref{domega=0} and so it is indeed strictly topological. 

Finding $\Psi_t$, $S_{\mathrm{top}}$ when $H\not=0$ is far more difficult.
In this case, apparently, the target space tensor fields which can be built directly from 
$g$, $H$, $K_\pm$ are not sufficient for constructing a gauge fermion $\Psi_t$ and 
a topological action $S_\mathrm{top}$. 
So far, we have not been able to find the solution of this problem in
full generality. We have however found a solution valid in the generic
situation, as we illustrate next.

Below, we shall assume that the pure spinor $\phi_1$ of the $H$ twisted 
generalized complex structure $\mathcal{J}_1$ associated with by the biHermitian 
structure $(g,H,K_\pm)$ via \eqref{J1/2} can be taken of the form 
\begin{equation}\label{phi'1}
\phi_1=\exp_\wedge(b+\beta), 
\end{equation}
where $\beta$ is a complex $2$--form. In our case, for reasons explained in
sect. \ref{sec:(2,2)susysigma}, $\mathcal{J}_1$ is actually a $H$ twisted weak generalized
Calabi--Yau structure and, so, the pure spinor $\phi_1$ is globally defined
and satisfies \eqref{weakCY}. This requires that $\beta$ is closed,
\begin{equation}\label{dbeta1=0}
d\beta=0. 
\end{equation}
Twisted generalized complex structures satisfying \eqref{phi'1} are generic, since, in a
sense, most of them do, as shown in refs. \cite{Hitchin1, Gualtieri}.
Generalized Kaehler structures with the above properties have
been considered by Hitchin in \cite{Hitchin2}, where various non trivial
examples are worked out in detail. 

Now, using \eqref{J1/2}, one verifies that the sections $X+\xi$ of 
$(TM \oplus T^*M)\otimes \mathbb{C}$ of the form \hphantom{xxxxxxxxxxxxxxxxxx}
\pagebreak[2]
\begin{equation}\label{eigenJ1}
X+\xi=X\mp igK_\pm X,
\end{equation}
with $X$ a section of $T^{10}_\pm M$ are valued in the $+i$ eigenbundle of 
$\mathcal{J}_1$. Thus, as explained in sect. \ref{sec:biHermitian}, 
these must annihilate the pure spinor $\phi_1$ (cf. eq. \eqref{Cliffact}). 
It is easy to see that this leads to the equation
\begin{equation}\label{iX(etc)=0}
i_X(\beta+b\pm igK_\pm)=0,
\end{equation}
for any section $X$ of $T^{10}_\pm M$. From here, it follows that there are 
two $2$--forms $\gamma_\pm$ of type $(2,0)$ with respect to
the complex structure $K_\pm$, respectively, such that 
\begin{equation}\label{defgamma1pm}
\beta+b\pm igK_\pm-\overline{\gamma}_\pm=0.
\end{equation}
This is our basic technical result.

The $2$--forms $\gamma_\pm$ furnish the hitherto missing elements needed
for the construction of the topological gauge fermion $\Psi_t$ and the
topological action $S_\mathrm{top}$. The crucial relations leading to their existence 
and determining their properties are \eqref{dbeta1=0}, \eqref{iX(etc)=0},
which however hinge on the assumption that the pure spinor $\phi_1$ is of the
form \eqref{phi'1}. There are of course biHermitian structures for which \eqref{phi'1}, 
is not fulfilled. In general, the pure spinor $\phi_1$ is of
the form
\begin{equation}\label{phi1gen}
\phi_1=\exp_\wedge(b+\beta)\wedge\Omega, 
\end{equation}
where $\beta$ is a complex $2$--form and $\Omega$ is a complex 
$k$--form that is decomposable 
\begin{equation}\label{Omegafact}
\Omega=\theta_1\wedge\cdots\wedge\theta_k, 
\end{equation}
the $\theta_i$ being linearly independent 
complex $1$--forms \cite{Hitchin1, Gualtieri}.
The integer $k$ is called type. 
Demanding that $\phi_1$ satisfies the twisted weak Calabi--Yau
condition \eqref{weakCY} entails the equations \hphantom{xxxxxxxxxxxxx}
\begin{subequations}\label{weakCYgen}
\begin{align}
&d\Omega=0,
\label{dOmega=0}\\
&d\beta\wedge\Omega=0.
\label{dbetaOmega=0}
\end{align}
\end{subequations}
Requiring further that sections $X+\xi$ of 
$(TM \oplus T^*M)\otimes \mathbb{C}$ of the form \eqref{eigenJ1}
annihilate $\phi_1$ yields 
\vskip-1cm
\begin{subequations}\label{iX}
\begin{align}
&i_X\Omega=0,
\label{iXOmega=0}\\
&i_X(\beta+b\pm igK_\pm)\wedge\Omega=0.
\label{iX(etc)Omega=0}
\end{align}
\end{subequations}
for any section $X$ of $T^{10}_\pm M$. In this way, 
we see that, while  \eqref{dbeta1=0}, \eqref{iX(etc)=0} hold in the generic case
\eqref{phi'1}, they do not necessarily hold in the non generic case \eqref{phi1gen}, 
though they may. If they do, then $2$--forms $\gamma_\pm$ exist and have the
same properties as in the generic case. 

The type $k$ is not necessarily constant and may jump at a locus $C\subset M$ 
of an even number of units. Type jumping is one of the subtlest aspects of
generalized complex geometry \cite{Gualtieri,Cavalcanti}. If it does occur, it is possible for the spinor 
$\phi_1$ to have the generic form \eqref{phi'1} at $M\setminus C$, while
taking the non generic form \eqref{phi1gen} at $C$. In that 
case, we expect the $2$--forms $\gamma_\pm$ to develop some sort of
singularity at $C$. If the embedding field $x$ intersects $C$, 
then our analysis below, which assumes the smoothness of the $\gamma_\pm$,
may break down. In this way, the locus $C$ may behave as some kind of defect, 
that is invisible at the classical level, but which may have detectable 
effects at the quantum level. This however is just a speculation for the time being.
At any rate, type jumping occurs only for $\dim_\mathbb{R}M\geq 6$. Examples of type
jumping from $0$ to an higher even value are not easily found. 

Under the assumption that the $2$--forms $\gamma_\pm$ are available, one can show by 
explicit computation that \eqref{St=deltaPsi} indeed holds with
\begin{subequations}\label{PsiStopHnot=0}
\begin{align}
\Psi_t&=-\int_\Sigma d^2z\frac{1}{2}\Big\{
\big(g_{ab}+\frac{1}{2}\gamma_{+ab}\big)(x)
\psi_{+z}{}^a\overline{\partial}_{\overline{z}}x^b
\label{PsiHnot=0}\\
&\hskip6cm+\big(g_{ab}-\frac{1}{2}\gamma_{-ab}\big)(x)
\psi_{-\overline{z}}{}^a\partial_zx^b\Big\},
\nonumber\\
S_\mathrm{top}&=\int_\Sigma d^2z\frac{1}{4}\big(2b_{ab}
-iK_{+ab}+iK_{-ab}-\gamma_{+ab}-\gamma_{-ab}\big)(x)\partial_zx^a\overline{\partial}_{\overline{z}}x^b.
\label{StopHnot=0}
\end{align}
\end{subequations}
The verification requires the use of several non trivial identities involving 
$\gamma_\pm$ following from \eqref{dbeta1=0},
\eqref{defgamma1pm}, which are conveniently collected in appendix 
\ref{sec:appendixgammapm}. From \eqref{defgamma1pm}, it appears that 
the action $S_\mathrm{top}$ can be written as 
\pagebreak[2]
\begin{equation}\label{Stopbeta}
S_\mathrm{top}=-\int_\Sigma d^2z\frac{1}{2}\overline{\beta}_{ab}(x)
\partial_zx^a\overline{\partial}_{\overline{z}}x^b.
\end{equation}
Since $\beta$ satisfies \eqref{dbeta1=0}, the action $S_\mathrm{top}$ is again of the
form \eqref{Stopprop}, \eqref{domega=0} and, therefore, it is strictly topological. 
It is quite remarkable that $S_\mathrm{top}$ is related in simple fashion to the pure spinor
$\phi_1$ associated with the generalized complex structure $\mathcal{J}_1$
\footnote{\vphantom{$\Bigg]$} The possibility of a connection of this type was predicted by
A. Tomasiello before this analysis was made.}. 

In the above discussion, we have tacitly assume that the closed $3$--form
$H$ is exact, so that the $2$ form $b$ is globally defined. If 
$H$ is not exact, $b$ is defined only locally. The combination
$\beta+b$ is however globally defined in any case, as $\phi_1$ is, and, so, 
also the $2$--forms $\gamma_\pm$ are, by \eqref{defgamma1pm}. 
If $H$ is not exact, the meaning of the term 
$\int_\Sigma x^*b$ appearing in the expression of $S_\mathrm{top}$ must be
qualified. If $x(\Sigma)$ is a boundary in the target space
$M$,  then $\int_\Sigma x^*b=\int_\Gamma \bar{x}^* H$, 
where $\Gamma$ is a $3$-fold such that $\partial\Gamma=\Sigma$ 
and $\bar{x}:\Gamma\rightarrow M$ is an embedding such that $ \bar{x}|_\Sigma=x$.
The value of  $\int_\Sigma x^*b$ computed in this way depends on 
the choice of $\Gamma$. In the quantum theory, in order to have a 
well defined weight $\exp(iS)$ in the path integral for a properly normalized
action $S$, it is necessary to require that $H/2\pi$ has integer 
periods, so that the cohomology class $[H/2\pi]\in H^3(M,\mathbb{R})$ 
belongs to the image of $H^3(M,\mathbb{Z})$ in $H^3(M,\mathbb{R})$.
If one wants to extend the definition 
to the general case where $x(\Sigma)$ is a cycle of $M$, the theory of 
Cheeger--Simons differential characters is required \cite{Cheeger1,Cheeger2}. 

We remark that, when $H=0$, \eqref{St=deltaPsi} holds with $\Psi_t$, 
$S_\mathrm{top}$ given by \eqref{PsiH=0}, \eqref{StopH=0} even if
\eqref{phi'1} does not hold, i. e. the underlying twisted
generalized complex structure $\mathcal{J}_1$ is not generic. 
If it does, however, one can use alternatively 
\eqref{PsiHnot=0}, \eqref{StopHnot=0}. 

Assuming again that the $2$--forms $\gamma_\pm$ are available, 
one has also a chirally split version of \eqref{St=deltaPsi}, 
\begin{equation}\label{St=spmPsipm}
S_t\approx s_{t+}\Psi_{t+}+s_{t-}\Psi_{t-}+S_\mathrm{top},
\end{equation}
where the gauge fermions $\Psi_{t\pm}$ are given by 
\pagebreak[2]
\begin{subequations}\label{Psipmdef}
\begin{align}
\Psi_{t+}&=-\int_\Sigma d^2z\frac{1}{2}
\big(g_{ab}+\frac{1}{2}\gamma_{+ab}\big)(x)
\psi_{+z}{}^a\overline{\partial}_{\overline{z}}x^b,
\label{Psi+}\\
\Psi_{t-}&=-\int_\Sigma d^2z\frac{1}{2}\big(g_{ab}-\frac{1}{2}\gamma_{-ab}\big)(x)
\psi_{-\overline{z}}{}^a\partial_zx^b,
\label{Psi-}
\end{align}
\end{subequations}
and $S_\mathrm{top}$ is given by \eqref{StopHnot=0}. Note also that
\begin{equation}\label{Psi=Psi++Pis-}
\Psi_t=\Psi_{t+}+\Psi_{t-}
\end{equation}
When $H=0$, \eqref{St=spmPsipm} holds in any case with $\gamma_\pm=0$. 
The significance of these properties is not clear to us yet. 

The results, which we have obtained, albeit still incomplete, 
shed light on the nature of world sheet and target space geometrical data, 
on which the quantum field theories associated with the biHermitian $A$ and
$B$ sigma models effectively depend. 
The expressions \eqref{StopH=0}, \eqref{StopHnot=0} of $S_\mathrm{top}$ obtained above 
show that $S_\mathrm{top}$ depends only on $\mathcal{J}_1$
(cf. eq. \eqref{J1/2}).
Thus, the quantum biHermitian $B$ model considered here depends 
effectively only on $\mathcal{J}_1$. The quantum biHermitian $A$ model 
depends instead only on $\mathcal{J}_2$ on account of \eqref{ABcomp}.
Both models are also evidently independent from the complex structure of the  
world sheet $\Sigma$. These findings confirm earlier results 
\cite{Kapustin1,Kapustin2}.

\vfill\eject

\vskip.6cm\par\noindent{\bf Acknowledgments.}  
We thank A. Kapustin, V. Pestun and G. Cavalcanti,
for correspondence and G. Bonelli, 
A. Tanzini and A. Tomasiello for useful discussions. 
We warmly thank the Erwin Schroedinger Institute for Mathematical Physics 
(ESI) for the kind hospitality offered to us in June and July 2006.  
\vfill\eject

\appendix

\begin{small}
\section{  \bf Formulae of biHermitian geometry}
\label{sec:appendixtens}
\end{small}
In this appendix, we collect a number of useful identities of biHermitian
geometry, which are repeatedly used in the calculations illustrated in the
main body of the paper. Below $(g, H,K_\pm)$
is a fixed biHermitian structure on an even dimensional manifold $M$. 

\begin{enumerate}
\item
Relations satisfied by the 3--form $H_{abc}$.
\begin{equation}
\partial_aH_{bcd}-\partial_bH_{acd}+\partial_cH_{abd}-\partial_dH_{abc}=0.
\end{equation}
 
\item
Relations satisfied by the connections $\Gamma_\pm{}^a{}_{bc}$.
\begin{subequations}
\begin{align}
&\Gamma_\pm{}^a{}_{bc}=\Gamma^a{}_{bc}\pm\frac{1}{2}H^a{}_{bc},
\\
&\Gamma_\pm{}^a{}_{bc}=\Gamma_\mp{}^a{}_{cb},
\end{align}
\end{subequations}
where $\Gamma^a{}_{bc}$ is the Levi--Civita connection of the metric $g_{ab}$.

\item
Relations satisfied by the torsion $T_\pm{}^a{}_{bc}$ of $\Gamma_\pm{}^a{}_{bc}$.
\begin{subequations}
\begin{align}
&T_\pm{}^a{}_{bc}=\pm H^a{}_{bc},
\\
&T_\pm{}^a{}_{bc}=T_\mp{}^a{}_{cb}.
\end{align}
\end{subequations}
 
\item
Relations satisfied by the Riemann tensor $R_{\pm abcd}$ of $\Gamma_\pm{}^a{}_{bc}$.
\begin{subequations}
\begin{align}
&R_{\pm abcd}=R_{abcd}\pm\frac{1}{2}(\nabla_dH_{abc}-\nabla_cH_{abd})
+\frac{1}{4}(H^e{}_{ad}H_{ebc}-H^e{}_{ac}H_{ebd}),
\\
&R_{\pm abcd}=R_{\mp cdab},
\end{align}
\end{subequations}
where $R_{abcd}$ is the Riemann tensor of the metric $g_{ab}$.
\par
Bianchi identities. 
\begin{subequations}
\begin{align}
&R_{\pm abcd}+R_{\pm acdb}+R_{\pm adbc}
\mp(\nabla_{\pm b}H_{acd}+\nabla_{\pm c}H_{adb}+\nabla_{\pm d}H_{abc})
\\
&\hskip5cm +H^e{}_{ab}H_{ecd}+H^e{}_{ac}H_{edb}+H^e{}_{ad}H_{ebc}=0,
\nonumber\\
&\nabla_{\pm e}R_{\pm abcd}+\nabla_{\pm c}R_{\pm abde}+\nabla_{\pm d}R_{\pm abec}
\\
&\hskip3.5cm \pm(H^f{}_{ec}R_{\pm abfd}+H^f{}_{cd}R_{\pm abfe}+H^f{}_{de}R_{\pm abfc})=0.
\nonumber
\end{align}
\end{subequations}
Other identities 
\begin{subequations}
\begin{align}
&R_{\pm abcd}-R_{\pm cbad}=R_{\pm acbd}\pm\nabla_{\pm d}H_{abc},
\\
&R_{\pm abcd}-R_{\pm cbad}=R_{\mp acbd}\mp\nabla_{\mp b}H_{acd},
\\
&R_{\pm abcd}-R_{\mp abcd}=\pm\nabla_{\pm d}H_{abc}\mp\nabla_{\pm c}H_{dab}
\\
&\hskip4.5cm +H^e{}_{ac}H_{ebd}+H^e{}_{da}H_{ebc}-H^e{}_{ab}H_{ecd}.
\nonumber
\end{align}
\end{subequations}

\item
The complex structures $K_\pm{}^a{}_cK_\pm{}^c{}_b$.
\begin{equation}
K_\pm{}^a{}_cK_\pm{}^c{}_b=-\delta^a{}_b.
\end{equation}
Integrability
\begin{equation}
K_\pm{}^d{}_a\partial_dK_\pm{}^c{}_b-K_\pm{}^d{}_b\partial_dK_\pm{}^c{}_a
-K_\pm{}^c{}_d\partial_aK_\pm{}^d{}_b+K_\pm{}^c{}_d\partial_bK_\pm{}^d{}_a=0.
\end{equation}
Hermiticity \hphantom{xxxxxxxxxxxxxxxxx}
\begin{equation}
g_{cd}K_\pm{}^c{}_aK_\pm{}^d{}_b=g_{ab}.
\end{equation}
Kaehlerness with torsion \hphantom{xxxxxxxxxxxxxxxxx}
\begin{equation}
\nabla_{\pm a}K_\pm{}^b{}_c=0.
\end{equation}

\item 
Other properties.
\begin{equation}
H_{efg}\Lambda_\pm{}^e{}_a\Lambda_\pm{}^f{}_b\Lambda_\pm{}^g{}_c=0~~\text{and c. c.},
\end{equation}
\begin{equation}
R_{\pm efcd}\Lambda_\pm{}^e{}_a\Lambda_\pm{}^f{}_b=0~~\text{and c. c.},
\end{equation}
where \hphantom{xxxxxxxxxxxxxxxxx}
\begin{equation}
\Lambda_\pm{}^a{}_b=\frac{1}{2}(\delta^a{}_b-iK_\pm{}^a{}_b)~~\text{and c. c.}
\end{equation}

\end{enumerate}

\vfill\eject

\begin{small}
\section{  \bf Some technical calculations}
\label{sec:appendixsfeq}
\end{small}

Let $\mathscri{F}$ be the graded commutative algebra of local composite fields generated by the
fields $x^a$, $\overline{\chi}_+{}^a$, $\overline{\chi}_-{}^a$,
$\psi_{+z}{}^a$, $\psi_{+\overline{z}}{}^a$. Let $\mathscri{E}$ be the
bilateral ideal of $\mathscri{F}$ generated by the composite fields
\begin{subequations}\label{feid}
\begin{align}
D_{z\overline{z}}{}^a&=-\overline{\nabla}_{+\overline{z}}\partial_zx^a
+iR_{+bc}{}^a{}_d(x)\overline{\chi}_+{}^b\psi_{+z}{}^c\overline{\partial}_{\overline{z}}x^d
+iR_{-bc}{}^a{}_d(x)\overline{\chi}_-{}^b\psi_{-\overline{z}}{}^c\partial_zx^d
\\
&\phantom{=}+(\nabla_+{}^aR_{+bcde}+H^{fa}{}_dR_{+bcfe}+H^{fa}{}_eR_{+bcdf})(x)
\overline{\chi}_+{}^b\psi_{+z}{}^c\overline{\chi}_-{}^d\psi_{-\overline{z}}{}^e
\nonumber\\
&=-\nabla_{-z}\overline{\partial}_{\overline{z}}x^a
+iR_{+bc}{}^a{}_d(x)\overline{\chi}_+{}^b\psi_{+z}{}^c\overline{\partial}_{\overline{z}}x^d
+iR_{-bc}{}^a{}_d(x)\overline{\chi}_-{}^b\psi_{-\overline{z}}{}^c\partial_zx^d
\nonumber\\
&\phantom{=}+(\nabla_-{}^aR_{-bcde}-H^{fa}{}_dR_{-bcfe}-H^{fa}{}_eR_{-bcdf})(x)
\overline{\chi}_-{}^b\psi_{-\overline{z}}{}^c\overline{\chi}_+{}^d\psi_{+z}{}^e,
\nonumber\\
\overline{E}_{+\overline{z}}{}^a&=i\overline{\nabla}_{+\overline{z}}\overline{\chi}_+{}^a
+R_+{}^a{}_{bcd}(x)\overline{\chi}_+{}^b\overline{\chi}_-{}^c\psi_{-\overline{z}}{}^d,
\\
\overline{E}_{-z}{}^a&=i\nabla_{-z}\overline{\chi}_-{}^a
+R_-{}^a{}_{bcd}(x)\overline{\chi}_-{}^b\overline{\chi}_+{}^c\psi_{+z}{}^d,
\nonumber\\
F_{+\overline{z}z}{}^a&=i\overline{\nabla}_{+\overline{z}}\psi_{+z}{}^a
+R_+{}^a{}_{bcd}(x)\psi_{+z}{}^b\overline{\chi}_-{}^c\psi_{-\overline{z}}{}^d,
\\
F_{-z\overline{z}}{}^a&=i\nabla_{-z}\psi_{-\overline{z}}{}^a
+R_-{}^a{}_{bcd}(x)\psi_{-\overline{z}}{}^b\overline{\chi}_+{}^c\psi_{+z}{}^d.
\nonumber
\end{align}
\end{subequations}
$\mathscri{E}$ is usually called the ideal of field equations, because the 
vanishing of its generators \eqref{feid} is equivalent to the imposition of
the field equations on the basic fields. The on shell quotient algebra
$\mathscri{F}_\mathscri{E}=\mathscri{F}/\mathscri{E}$ is thus defined. 

The ideal $\mathscri{E}$ is invariant under the fermionic variation operators 
$s_{t+}$, $s_{t-}$,  defined in \eqref{stoppm}, as the following calculation shows
\begin{subequations}
\begin{align}
&s_{t+}D_{z\overline{z}}{}^a=-i\Gamma_-{}^a{}_{cb}(x)\overline{\chi}_+{}^cD_{z\overline{z}}{}^b
-\nabla_{-z}\overline{E}_{+\overline{z}}{}^a+iR_-{}^a{}_{dbc}(x)\overline{E}_{+\overline{z}}{}^d
\overline{\chi}_+{}^b\psi_{+z}{}^c
\\
&\hskip1.9cm 
-iR_+{}^a{}_{bcd}(x)\overline{\chi}_+{}^b\overline{E}_{-z}{}^c\psi_{-\overline{z}}{}^d
-iR_+{}^a{}_{bcd}(x)\overline{\chi}_+{}^b\overline{\chi}_-{}^cF_{-z\overline{z}}{}^d,
\nonumber\\
&s_{t-}D_{z\overline{z}}{}^a=-i\Gamma_+{}^a{}_{cb}(x)\overline{\chi}_-{}^cD_{z\overline{z}}{}^b
-\overline{\nabla}_{+\overline{z}}\overline{E}_{-z}{}^a+iR_+{}^a{}_{dbc}(x)\overline{E}_{-z}{}^d
\overline{\chi}_-{}^b\psi_{-\overline{z}}{}^c
\nonumber\\
&\hskip1.9cm 
-iR_-{}^a{}_{bcd}(x)\overline{\chi}_-{}^b\overline{E}_{+\overline{z}}{}^c\psi_{+z}{}^d
-iR_-{}^a{}_{bcd}(x)\overline{\chi}_-{}^b\overline{\chi}_+{}^cF_{+\overline{z}z}{}^d,
\nonumber\\
&s_{t+}\overline{E}_{+\overline{z}}{}^a=-i\Gamma_-{}^a{}_{cb}(x)\overline{\chi}_+{}^c
\overline{E}_{+\overline{z}}{}^b,
\\
&s_{t-}\overline{E}_{+\overline{z}}{}^a=-i\Gamma_+{}^a{}_{cb}(x)\overline{\chi}_-{}^c
\overline{E}_{+\overline{z}}{}^b,
\nonumber\\
&s_{t+}\overline{E}_{-z}{}^a=-i\Gamma_-{}^a{}_{cb}(x)\overline{\chi}_+{}^c\overline{E}_{-z}{}^b,
\\
&s_{t-}\overline{E}_{-z}{}^a=-i\Gamma_+{}^a{}_{cb}(x)\overline{\chi}_-{}^c\overline{E}_{-z}{}^b,
\nonumber
\end{align}
\begin{align}
&s_{t+}F_{+\overline{z}z}{}^a=-i\Gamma_+{}^a{}_{cb}(x)\overline{\chi}_+{}^cF_{+\overline{z}z}{}^b
\\
&\hskip1.9cm +i\Lambda_+{}^a{}_b(x)[D_{z\overline{z}}{}^b
+H^b{}_{cd}(x)\overline{E}_{+\overline{z}}{}^c\psi_{+z}{}^d
+H^b{}_{cd}(x)\overline{\chi}_+{}^cF_{+\overline{z}z}{}^d],
\nonumber\\
&s_{t-}F_{+\overline{z}z}{}^a=-i\Gamma_+{}^a{}_{cb}(x)\overline{\chi}_-{}^cF_{+\overline{z}z}{}^b,
\nonumber\\
&s_{t+}F_{-z\overline{z}}{}^a=-i\Gamma_-{}^a{}_{cb}(x)\overline{\chi}_+{}^cF_{-z\overline{z}}{}^b,
\\
&s_{t-}F_{-z\overline{z}}{}^a=-i\Gamma_-{}^a{}_{cb}(x)\overline{\chi}_-{}^cF_{-z\overline{z}}{}^b
\nonumber\\
&\hskip1.9cm +i\Lambda_-{}^a{}_b(x)[D_{z\overline{z}}{}^b-H^b{}_{cd}(x)
\overline{E}_{-z}{}^c\psi_{-\overline{z}}{}^d
-H^b{}_{cd}(x)\overline{\chi}_-{}^cF_{-z\overline{z}}{}^d].
\nonumber
\end{align}
\end{subequations}
Therefore, $s_{t+}$, $s_{t-}$ induce fermionic variation operators on the on shell algebra
$\mathscri{F}_\mathscri{E}$, which we shall denote by the same symbols. 
The composite variations $s_{t+}{}^2$, $s_{t-}{}^2$, 
$s_{t+}s_{t-}+s_{t-}s_{t+}$map the field algebra $\mathscri{F}$ into the field
equation ideal $\mathscri{E}$, as 
\begin{subequations}
\begin{align}
&s_{t+}{}^2x^a=0,
\\
&s_{t-}{}^2x^a=0,
\nonumber\\
&(s_{t+}s_{t-}+s_{t-}s_{t+})x^a=0,
\nonumber\\
&s_{t+}{}^2\overline{\chi}_+{}^a=0,
\\
&s_{t-}{}^2\overline{\chi}_+{}^a=0,
\nonumber\\
&(s_{t+}s_{t-}+s_{t-}s_{t+})\overline{\chi}_+{}^a=0,
\nonumber\\
&s_{t+}{}^2\overline{\chi}_-{}^a=0,
\\
&s_{t-}{}^2\overline{\chi}_-{}^a=0,
\nonumber\\
&(s_{t+}s_{t-}+s_{t-}s_{t+})\overline{\chi}_-{}^a=0,
\nonumber\\
&s_{t+}{}^2\psi_{+z}{}^a=0,
\\
&s_{t-}{}^2\psi_{+z}{}^a=0,
\nonumber\\
&(s_{t+}s_{t-}+s_{t-}s_{t+})\psi_{+z}{}^a=-\Lambda_+{}^a{}_b(x)\overline{E}_{-z}{}^b,
\nonumber\\
&s_{t+}{}^2\psi_{-\overline{z}}{}^a=0,
\\
&s_{t-}{}^2\psi_{-\overline{z}}{}^a=0,
\nonumber\\
&(s_{t+}s_{t-}+s_{t-}s_{t+})\psi_{-\overline{z}}{}^a=-\Lambda_-{}^a{}_b(x)\overline{E}_{+\overline{z}}{}^b.
\nonumber
\end{align}
\end{subequations}
Therefore, $s_{t+}$, $s_{t-}$ are nilpotent and anticommute on
$\mathscri{F}_\mathscri{E}$. This shows \eqref{stoppm2=0}.

Instead of the field algebra $\mathscri{F}$, we consider now the graded commutative 
form field algebra $\mathscri{F}^\bullet$ generated by the scalar fields
$x^a$, $\overline{\chi}_+{}^a$, $\overline{\chi}_-{}^a$, $\psi_+{}^a$,
$\psi_-{}^a$, where $\psi_+{}^a$, $\psi_-{}^a$ are defined in \eqref{formpsi}. 
Likewise, we consider the bilateral ideal $\mathscri{E}^\bullet$ of
$\mathscri{F}^\bullet$ generated by the form field equation fields 
\begin{subequations}
\begin{align}
&D^a=dz\wedge d\overline{z}D_{z\overline{z}}{}^a,
\\
&\overline{E}_+{}^a=d\overline{z}\overline{E}_{+\overline{z}}{}^a,
\qquad \hskip.9cm \overline{E}_-{}^a=dz \overline{E}_{-z}{}^a,
\\
&F_+{}^a=d\overline{z}\wedge dz F_{+\overline{z}z}{}^a,
\qquad F_-{}^a=dz\wedge d\overline{z}F_{-z\overline{z}}{}^a,
\end{align}
\end{subequations}
where $D_{z\overline{z}}{}^a$, $\overline{E}_{+\overline{z}}{}^a$, $\overline{E}_{-z}{}^a$, $
F_{+\overline{z}z}{}^a$, $F_{-z\overline{z}}{}^a$ are defined in \eqref{feid}. 
The on shell form field algebra $\mathscri{F}^\bullet{}_{\mathscri{E}^\bullet}=
\mathscri{F}^\bullet/\mathscri{E}^\bullet$ is therefore defined. 

The fermionic variation operators $s_{t+}$, $s_{t-}$ extend in natural and obvious fashion 
to the field algebras $\mathscri{F}^\bullet{}$ and
$\mathscri{F}^\bullet{}_{\mathscri{E}^\bullet}$, upon assuming
conventionally that $s_{t+}$, $s_{t-}$ anticommute with $dz$,
$d\overline{z}$. 
In addition to $s_{t+}$, $s_{t-}$, we have also the 
even 1--form variations $h_{t+}$, $h_{t-}$, defined by eqs. \eqref{htoppm},
\eqref{htoppm1f} and acting on $\mathscri{F}^\bullet$. $h_{t+}$, $h_{t-}$
preserve $\mathscri{E}^\bullet$, since indeed
\begin{subequations}
\begin{align}
&h_{t+}D^a=0,
\\
&h_{t-}D^a=0,
\nonumber\\
&h_{t+}\overline{E}_+{}^a=-i\Gamma_+{}^a{}_{cb}(x)\psi_+{}^c\wedge \overline{E}_+{}^b
\\
&\hskip1.9cm +i\overline{\Lambda}_+{}^a{}_b(x)[-D^b
+H^b{}_{cd}(x)\psi_+{}^c \wedge \overline{E}_+{}^d
+H^b{}_{cd}(x)\overline{\chi}_+{}^cF_+{}^d],
\nonumber\\
&h_{t-}\overline{E}_+{}^a=0,
\nonumber\\
&h_{t+}\overline{E}_-{}^a=0,
\\
&h_{t-}\overline{E}_-{}^a=-i\Gamma_-{}^a{}_{cb}(x)\psi_-{}^c\wedge \overline{E}_-{}^b
\nonumber\\
&\hskip1.9cm +i\overline{\Lambda}_-{}^a{}_b(x)[D^b
-H^b{}_{cd}(x)\psi_-{}^c \wedge \overline{E}_-{}^d
-H^b{}_{cd}(x)\overline{\chi}_-{}^cF_-{}^d],
\nonumber\\
&h_{t+}F_+{}^a=0,
\\
&h_{t-}F_+{}^a=0,
\nonumber\\
&h_{t+}F_-{}^a=0,
\\
&h_{t-}F_-{}^a=0.
\nonumber
\end{align}
\end{subequations}
We note that the relations $dz\wedge dz=0$, $d\overline{z}\wedge
d\overline{z}=0$ are crucial for ensuring the validity of the above algebra.
Therefore, $h_{t+}$, $h_{t-}$ induce even 1--form variations on the on shell
form field algebra
$\mathscri{F}^\bullet{}_{\mathscri{E}^\bullet}$, which we shall denote by the same symbols. 
An explicit calculation using \eqref{stoppm}, \eqref{htoppm},
\eqref{htoppm1f} on the same lines as the above yields the relations 
\begin{subequations}
\begin{align}
&(h_{t+}s_{t+}-s_{t+}h_{t+})x^a=-i\partial x^a,
\\
&(h_{t-}s_{t-}-s_{t-}h_{t-})x^a=-i\overline{\partial} x^a,
\nonumber\\
&(h_{t+}s_{t-}-s_{t-}h_{t+})x^a=0,
\nonumber\\
&(h_{t-}s_{t+}-s_{t+}h_{t-})x^a=0,
\nonumber\\
&(h_{t+}s_{t+}-s_{t+}h_{t+})\overline{\chi}_+{}^a=-i\partial \overline{\chi}_+{}^a,
\\
&(h_{t-}s_{t-}-s_{t-}h_{t-})\overline{\chi}_+{}^a=-i\overline{\partial} \overline{\chi}_+{}^a+\overline{E}_+{}^a,
\nonumber\\
&(h_{t+}s_{t-}-s_{t-}h_{t+})\overline{\chi}_+{}^a=-\overline{\Lambda}_+{}^a{}_b(x)\overline{E}_-{}^b,
\nonumber\\
&(h_{t-}s_{t+}-s_{t+}h_{t-})\overline{\chi}_+{}^a=0,
\nonumber\\
&(h_{t+}s_{t+}-s_{t+}h_{t+})\overline{\chi}_-{}^a=-i\partial\overline{\chi}_-{}^a+\overline{E}_-{}^a,
\\
&(h_{t-}s_{t-}-s_{t-}h_{t-})\overline{\chi}_-{}^a=-i\overline{\partial}\overline{\chi}_-{}^a,
\nonumber\\
&(h_{t+}s_{t-}-s_{t-}h_{t+})\overline{\chi}_-{}^a=0,
\nonumber\\
&(h_{t-}s_{t+}-s_{t+}h_{t-})\overline{\chi}_-{}^a=
-\overline{\Lambda}_-{}^a{}_b(x)\overline{E}_+{}^b,
\nonumber\\
&(h_{t+}s_{t+}-s_{t+}h_{t+})\psi_+{}^a=-i\partial\psi_+{}^a,
\\
&(h_{t-}s_{t-}-s_{t-}h_{t-})\psi_+{}^a=-i\overline{\partial}\psi_+{}^a+F_+{}^a,
\nonumber\\
&(h_{t+}s_{t-}-s_{t-}h_{t+})\psi_+{}^a=0,
\nonumber\\
&(h_{t-}s_{t+}-s_{t+}h_{t-})\psi_+{}^a=\Lambda_+{}^a{}_b(x)F_-{}^b,
\nonumber\\
&(h_{t+}s_{t+}-s_{t+}h_{t+})\psi_-{}^a=-i\partial \psi_-{}^a,
+F_-{}^a
\\
&(h_{t-}s_{t-}-s_{t-}h_{t-})\psi_-{}^a=-i\overline{\partial}\psi_-{}^a ,
\nonumber\\,
&(h_{t+}s_{t-}-s_{t-}h_{t+})\psi_-{}^a=\Lambda_-{}^a{}_b(x)F_+{}^b,
\nonumber\\
&(h_{t-}s_{t+}-s_{t+}h_{t-})\psi_-{}^a=0,
\nonumber
\end{align}
\end{subequations}
and, similarly
\pagebreak
\begin{subequations}
\begin{align}
&h_{t+}h_{t+}x^a=0,
\\
&h_{t-}h_{t-}x^a=0,
\nonumber
\\
&(h_{t+}h_{t-}-h_{t-}h_{t+})x^a=0,
\nonumber
\\
&h_{t+}h_{t+}\overline{\chi}_+{}^a=0,
\\
&h_{t-}h_{t-}\overline{\chi}_+{}^a=0,
\nonumber\\
&(h_{t+}h_{t-}-h_{t-}h_{t+})\overline{\chi}_+{}^a=
\overline{\Lambda}_+{}^a{}_b(x)F_-{}^b ,
\nonumber\\
&h_{t+}h_{t+}\overline{\chi}_-{}^a=0,
\\
&h_{t-}h_{t-}\overline{\chi}_-{}^a=0,
\nonumber\\
&(h_{t+}h_{t-}-h_{t-}h_{t+})\overline{\chi}_-{}^a=
-\overline{\Lambda}_-{}^a{}_b(x)F_+{}^b,
\nonumber\\
&h_{t+}h_{t+}\psi_+{}^a=0,
\\
&h_{t-}h_{t-}\psi_+{}^a=0,
\nonumber\\
&(h_{t+}h_{t-}-h_{t-}h_{t+})\psi_+{}^a=0,
\nonumber\\
&h_{t+}h_{t+}\psi_-{}^a=0,
\\
&h_{t-}h_{t-}\psi_-{}^a=0,
\nonumber\\
&(h_{t+}h_{t-}-h_{t-}h_{t+})\psi_-{}^a=0.
\nonumber
\end{align}
\end{subequations}
From these relation \eqref{hspm1fos}, \eqref{hhpm1fos} 
follow immediately.

Similar results hold for the BRST variation $s_t$ and the operator $h_t$, as is obvious from
\eqref{st=st++st-} and \eqref{ht}, respectively.

\vfill\eject

\begin{small}
\section{  \bf Relevant identities involving $\gamma_\pm$}
\label{sec:appendixgammapm}
\end{small}

To begin with, we note that, since $\gamma_\pm$ is a $2$--form of type $(2,0)$
with respect to $K_\pm$, one has
\begin{equation}
\overline{\Lambda}_\pm{}^c{}_a\gamma_{\pm cb}=0.
\end{equation}
This relation will be exploited throughout.

From \eqref{defgamma1pm}, it follows that 
\begin{equation}
H\mp id(gK_\pm)-d\gamma_\pm=0.
\end{equation}
This identity can be cast as
\begin{align}
\nabla_{\mp a}(\mp iK_{\pm bc}-\gamma_{\pm bc})
&=\nabla_{\pm b}\gamma_{\pm ca}+\nabla_{\pm c}\gamma_{\pm ab}\pm H^g{}_{bc}\gamma_{\pm ga}
-2\Lambda_\pm{}^d{}_aH_{dbc},
\end{align}
from which one obtains easily
\begin{subequations}\label{relevgamma}
\begin{align}
&\Lambda_\pm{}^d{}_a\nabla_{\mp d}(\mp iK_{\pm bc}-\gamma_{\pm bc})
\\
&\hskip2cm=\Lambda_\pm{}^d{}_a(\nabla_{\pm b}\gamma_{\pm cd}+\nabla_{\pm c}\gamma_{\pm db}
\pm H^g{}_{bc}\gamma_{\pm gd}-2H_{dbc}),
\nonumber\\
&\overline{\Lambda}_\pm{}^d{}_a\nabla_{\mp d}(\mp iK_{\pm bc}-\gamma_{\pm bc})=0.
\end{align}
\end{subequations}

From \eqref{defgamma1pm}, it follows also that 
\begin{equation}
igK_++igK_-+\gamma_+-\gamma_-=0.
\end{equation}
From here, one obtains that 
\begin{equation}\label{relevgamma1}
\overline{\Lambda}_\mp{}^f{}_c\nabla_{\mp a}(\mp iK_{\pm bf}-\gamma_{\pm bf})=0.
\end{equation}
Using \eqref{relevgamma}, \eqref{relevgamma1}, it is straightforward to verify
that \eqref{St=deltaPsi} holds with $\Psi_t$, $S_\mathrm{top}$ given by 
\eqref{PsiStopHnot=0}.

\vfill\eject

\end{document}